%% file: bdrive.tex
\documentclass[
 reprint,
superscriptaddress,
 amsmath,amssymb,
 aps,
 prx,
 longbibliography,
]{revtex4-2}
\usepackage{CJK}
\usepackage{graphicx}
\usepackage{dcolumn}
\usepackage{amssymb}
\usepackage{amsbsy}
\usepackage{amsmath}
\usepackage{mathrsfs}
\usepackage{epsfig}
\usepackage{graphicx}
\usepackage{array}
\usepackage{textcomp}
\usepackage{color}
\usepackage{braket}
\usepackage[unicode=true,pdfusetitle,
 bookmarks=true,bookmarksnumbered=false,bookmarksopen=false,
 breaklinks=false,pdfborder={0 0 0},pdfborderstyle={},backref=false,colorlinks=true]
 {hyperref}

\usepackage{comment}

\hypersetup{
 linkcolor=blue, urlcolor=blue, citecolor=blue}
 
\usepackage[justification=raggedright,font=small]{caption}
\usepackage[titletoc,toc,title]{appendix}
\usepackage[normalem]{ulem}
\usepackage[labelformat=simple]{subcaption}

\usepackage{tikz}

\usepackage{pgfplots}
\pgfplotsset{compat=1.18}
\usetikzlibrary{decorations.pathmorphing, arrows.meta}
\usetikzlibrary{plotmarks}

\usepackage{cleveref}
\crefname{appendix}{App.}{Apps.}
\crefname{equation}{Eq.}{Eqs.}
\crefname{figure}{Fig.}{Figs.}
\crefname{table}{Tab.}{Tabs.}
\crefname{section}{Sec.}{Secs.}

\newcommand{\beq}{\begin{equation}}
\newcommand{\eeq}{\end{equation}}

\newcommand{\bpm}{\begin{pmatrix}}
\newcommand{\epm}{\end{pmatrix}}
\newcommand{\bmm}{\begin{matrix}}
\newcommand{\emm}{\end{matrix}}

\newcommand{\Tr}{\mathrm{Tr}}

\def\be{\begin{equation}}
\def\ee{\end{equation}}
\def\bea{\begin{eqnarray}}
\def\eea{\end{eqnarray}}
\def\ie{\begin{equation}\begin{aligned}}
\def\fe{\end{aligned}\end{equation}}

\begin{document}
\title{Non-Equilibrium Phase Transition in a Boundary-Driven Dissipative Fermionic Chain}

\author{Hao Chen}
\thanks{These authors contributed equally to this work.}
\affiliation{
Department of Physics,
Princeton University, NJ 08544, USA
}
\affiliation{
Department of Electrical and Computer Engineering,
Princeton University, NJ 08544, USA
}

\author{Wucheng Zhang}
\thanks{These authors contributed equally to this work.}
\affiliation{
Department of Physics,
Princeton University, NJ 08544, USA
}

\author{Manas Kulkarni}
\affiliation{International Centre for Theoretical Sciences, Tata Institute of Fundamental Research, Bangalore 560089, India}

\author{Abhinav Prem}
\affiliation{Physics Program, Bard College, 30 Campus Road, Annandale-on-Hudson, New York 12504, USA}

\date{\today}

\begin{abstract}
We demonstrate that a boundary-localized periodic (Floquet) drive can induce nontrivial long-range correlations in a non-interacting fermionic chain which is additionally subject to boundary dissipation. Surprisingly, we find that this phenomenon occurs even when the corresponding isolated bulk is in a trivial gapped phase with exponentially decaying correlations. We argue that this boundary-drive induced non-equilibrium transition (as witnessed through the correlation matrix) is driven by a resonance mechanism whereby the drive frequency bridges bulk energy gaps, allowing boundary-injected particles and holes to propagate and mediate long-range correlations into the bulk. We also numerically establish that when the drive bridges a particle-hole gap, the induced long-range order scales as a power law with the bulk pairing potential ($\chi \sim \gamma^2$). Our results highlight the potential of localized coherent driving for generating macroscopic order in open quantum systems.

\end{abstract}

\maketitle

\paragraph*{Introduction---} The study of non-equilibrium steady-states (NESS) in open quantum systems has revealed that the interplay between incoherent dissipation and coherent driving can generate entangled quantum states and stabilize non-trivial quantum phases~\cite{mori2023,sato2025}. While the engineering of bulk driving and dissipation has been successfully used to prepare targeted many-body states~\cite{diehl2008,verstraete2009,fossfeig2017,mi2024,schnell2024,chirame2025,chirame2025b,chen2025eng,carusotto2025}, a more minimal and experimentally relevant paradigm involves systems driven and coupled to baths solely at their boundaries~\cite{PhysRevX.7.011016,fedorov2021}. Indeed, the possibility of coherently controlling macroscopic order via precisely engineering local boundary couplings has moved from abstract theory to experimental viability given rapid progress in controllable quantum platforms.

Intuitively, one might expect that the influence of a local bath and a coherent drive acting on the edge of a large system would decay exponentially into the bulk (unless the static bulk is already critical), leading to a featureless state in the thermodynamic limit. However, local coupling to a dissipative environment can indeed drive the emergence of non-trivial long-range order (LRO) in an initially short-range correlated bulk~\cite{prosen2010,prosen2010long,prosen2011,PhysRevLett.101.105701,parmee2020,dutta2021,wanisch2026} and even induce arrested relaxation~\cite{disszeno,wang2025dynamical}. In contrast, whether the interplay between a spatially localized monochromatic drive (which is far easier to experimentally engineer) and generic boundary dissipation can induce long-range correlations in an otherwise featureless bulk remains much less explored (see however Refs.~\cite{prem2023,ferrari2025,boundarydrivedoublechain,yao2025hidden} for related work.)

In this work, we demonstrate that a boundary-localized Floquet drive can indeed stabilize macroscopic order deep in the bulk. We find that even when the static bulk is in a trivial gapped phase with exponentially decaying correlations, the boundary drive triggers a non-equilibrium phase transition into a NESS with long-range correlations. We identify this transition as being mediated by a resonance mechanism: the drive frequency bridges the bulk energy gaps, effectively ``opening" channels for boundary-injected particles and holes to propagate and mediate correlations deep into the bulk. This extends the study of boundary-driven phase transitions to the Floquet domain, offering a new pathway for coherent control in locally driven-dissipative quantum systems.

\begin{figure}[t]
    \centering
    \input{fig1}
\caption{Schematic representation of the model: two fermionic chains (labeled by $\sigma = \uparrow, \downarrow$), each with hopping, pairing, and uniform on-site energy ($t_\sigma, \gamma_\sigma, h_\sigma$) are coupled at the first site by a monochromatic drive $H_D(t)$ with frequency $\omega$. The system is connected to Markovian baths at its boundaries (sites $j=1$ and $j=L$) with gain and loss rates given by $\Gamma_{j,\sigma,g}$ and $\Gamma_{j,\sigma,l}$, respectively.}
\label{fig:model_schematic}
\end{figure}
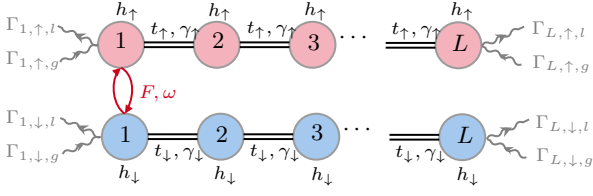

\paragraph*{Model---} We consider a system of two fermionic chains, labeled by spin indices $\sigma \in \{\uparrow, \downarrow\}$. The chains are coupled via a coherent monochromatic drive $H_D(t)$ locally acting on their first sites, with the total time-dependent Hamiltonian given by
\begin{equation} 
    H(t) = H_\uparrow + H_\downarrow + H_D(t)\,.
\end{equation}
See \cref{fig:model_schematic} for a schematic illustration. Each individual chain represents a Kitaev chain~\cite{kitaevchain} of length $L$ with open boundary conditions. The static Hamiltonian reads
\begin{equation} \label{eq:bulkH}
    \begin{aligned}
    H_\sigma = & \sum_{j=1}^{L-1} \left( t_\sigma c^{\dagger}_{j+1,\sigma}c_{j,\sigma} + \gamma_\sigma c^{\dagger}_{j+1,\sigma}c_{j,\sigma}^\dagger + \text{h.c.} \right) \\
    & + \sum_{j=1}^L h_\sigma (2c_{j,\sigma}^\dagger c_{j,\sigma}-1)\,,
    \end{aligned}
\end{equation}
where $c_{j,\sigma}^{\dagger}$ ($c_{j,\sigma}$) are fermionic creation (annihilation) operators satisfying $\{c_{j,\sigma}, c_{i,\sigma'}^\dagger\} = \delta_{ji}\delta_{\sigma\sigma'}$. The parameters $t_\sigma$, $\gamma_\sigma$, and $h_\sigma$ correspond to the hopping amplitude, superconducting pairing potential, and uniform on-site energy, respectively.

The static Hamiltonian above yields the bulk energy spectrum (i.e., the spectrum under the periodic boundary condition: $c_{L+1,\sigma} = c_{1,\sigma}$) consists of two energy bands
\begin{equation} \label{eq:sepctrum}
    E_{k,\sigma} = \pm 2 \sqrt{(h_\sigma+t_\sigma\cos k)^2+\gamma_\sigma^2\sin^2k}\,.
\end{equation}
The model Eq.~\eqref{eq:bulkH} is equivalent to the spin-$\frac{1}{2}$ XY-chain by the Jordan-Wigner transformation.

To facilitate the analytical treatment of the driven system, we impose a specific symmetry between the two chains
\begin{equation}
    t_\uparrow = - t_\downarrow\,, \quad \gamma_\uparrow = -\gamma_\downarrow\,, \quad h_\uparrow = - h_\downarrow\,.
    \label{eq:parameter_sym}
\end{equation}
In what follows, we set $t_\uparrow=1$ and use $\gamma_\uparrow = \gamma$ and $h_\uparrow = h$ as our free parameters. The chains are coupled via a driving term $H_D(t)$ acting on their first sites, which mixes the two spin species with frequency $\omega$:
\begin{equation}
    \begin{aligned}
        H_D(t) &= F \left[ \cos(\omega t) (c_{1,\uparrow}^\dagger c_{1,\uparrow} - c_{1,\downarrow} c_{1,\downarrow}^\dagger ) \right. \\
        &\quad \left. + \sin(\omega t) (c_{1,\uparrow}^\dagger c_{1,\downarrow}^\dagger + c_{1,\downarrow}c_{1,\uparrow}) \right]\,.
    \end{aligned}
    \label{eq:driving_H}
\end{equation}
This describes a time-periodic modulation of (i) the on-site energy at the left boundary and (ii) an inter-chain pairing term localized at the boundary. This form of driving is chosen to allow the correlated injection of single fermions into each chain via the boundary drive (note that, besides being unphysical, driving of the form $(c+c^\dagger)\cos(\omega t)$ in a chain of spinless fermions would also render the problem analytically intractable~\cite{boundarydrivedoublechain}).

The system interacts with the environment through Markovian dissipation localized at the boundary sites. The time evolution of the system's density matrix $\rho(t)$ is governed by the Lindblad master equation
\begin{equation}
    \frac{d\rho}{dt} = \mathcal{L}(t)\rho \equiv -i[H(t), \rho] + \mathcal{D}\rho\,,
\end{equation}
where the Liouville superoperator $\mathcal{L}(t)$ inherits the time-periodicity of $H(t)$. The dissipator $\mathcal{D}\rho \equiv \sum_\mu \Gamma_\mu \left(2L_\mu \rho L_\mu^\dagger - \{L_\mu^\dagger L_\mu, \rho\}\right)$ incorporates all jump operators $L_\mu$, describing single-particle gain and loss arising from coupling to baths at the boundary sites. Specifically, the index $\mu$ encompasses the site $j\in \{1, L\}$, spin $\sigma \in \{\uparrow, \downarrow\}$, and process type $\nu \in \{g, l\}$ (gain and loss), with corresponding operators $L_{j,\sigma,g} = c_{j,\sigma}^\dagger$ and $L_{j,\sigma,l} = c_{j,\sigma}$. The dissipation rates satisfy the symmetry relations $\Gamma_{j,\uparrow,g} = \Gamma_{j,\downarrow,l},\  \Gamma_{j,\downarrow,g} = \Gamma_{j,\uparrow,l}$, ensuring the dissipative dynamics respects $SO(2)$ symmetry (see \cref{app:RWA}). In this paper, we set $\Gamma_{1,\uparrow,g} = 0.3$, $\Gamma_{1,\uparrow,l} = 0.5$, $\Gamma_{L,\uparrow,g} = 0.1$, and $\Gamma_{L,\uparrow,l} = 0.5$.

\paragraph*{Observables---}
Since the Hamiltonian is quadratic and the jump operators are linear in fermionic operators, the dynamics preserve Gaussian states~\cite{prosen2008}, i.e., the state of the system is fully characterized by the correlation matrix (the two-point correlation functions), which is defined as
\begin{equation}
    C^{\sigma \sigma'}_{jk}(t) = \braket{c_{j,\sigma}^\dagger c_{k,\sigma'}}_t = \mathrm{Tr}[\rho(t) c_{j,\sigma}^\dagger c_{k,\sigma'}]\,.
\end{equation}
In the long-time limit, the system settles into a \emph{time-dependent} non-equilibrium steady state (NESS) that evolves periodically with time $\rho(t+T) = \rho(t)$, where $T=2\pi/\omega$, called a Floquet-Lindblad steady state~\cite{Spectral,FLSS1,FLSS2,FLSS3}.

Here, we are primarily interested in time-averaged expectation values of observables evaluated in the Floquet-Lindblad NESS. As derived in \cref{app:RWA}, by transforming to a rotating frame where the effective Hamiltonian is time-independent, we can relate the lab-frame time-averaged correlations $\overline{\braket{c_{j,\sigma}^\dagger c_{k,\sigma'}}}$, denoted as $C_{jk}^{\sigma
\sigma'}$, to the time-independent steady-state correlations in the rotating frame 
$\braket{c_{j,\sigma}^\dagger c_{k,\sigma'}}_R$. We then follow the standard approach of third quantization~\cite{prosen2008,prosen2011,prosen2017} to solve for $\braket{c_{j,\sigma}^\dagger c_{k,\sigma'}}_R$.

The intra-chain correlations average to
\begin{subequations}
\begin{align}
    C_{jk}^{\uparrow\uparrow} & = \frac{1}{2}\left(\braket{ c_{j,\uparrow}^\dag c_{k,\uparrow}}_R - \braket{ c_{k,\downarrow}^\dag c_{j,\downarrow}}_R + \delta_{jk}\right)\,, \\
    C_{jk}^{\downarrow\downarrow} & = \frac{1}{2}\left(\braket{ c_{j,\downarrow}^\dag c_{k,\downarrow}}_R - \braket{ c_{k,\uparrow}^\dag c_{j,\uparrow}}_R + \delta_{jk}\right)\,.
\end{align}
\end{subequations}
The inter-chain correlations are
\begin{equation}
    C_{jk}^{\uparrow\downarrow} = \frac{1}{2}\left(\braket{ c_{j,\uparrow}^\dag c_{k,\downarrow}}_R - \braket{ c_{k,\uparrow}^\dag c_{j,\downarrow}}_R \right)\,.
\end{equation}
These relations allow us to evaluate the time-averaged correlation functions solely from the NESS solution of the time-independent Lindbladian in the rotating frame.

\paragraph*{Results---} We now analyze the NESS properties of the boundary-driven dissipative chain defined above. We focus specifically on the regime where the NESS of the system without drive would be in the trivial short-range correlated (gapped) phase of the Kitaev chain. In the absence of driving, it is well established that for $h > h_c = 1-\gamma^2$, the corresponding XY chain resides in a paramagnetic phase with exponentially decaying correlations~\cite{PhysRevLett.101.105701}.

Our main finding is that the periodic boundary drive can induce a phase transition, driving the system from the short-range correlated phase into one exhibiting long-range correlations. We characterize this transition through the structure of the correlation matrix, the phase diagram in the frequency-amplitude plane, and scaling analysis with respect to the pairing potential.

\textit{(a) Drive-Induced Long-Range Correlations---}
\begin{figure*}[t]
    \centering
    \includegraphics[width=\textwidth]{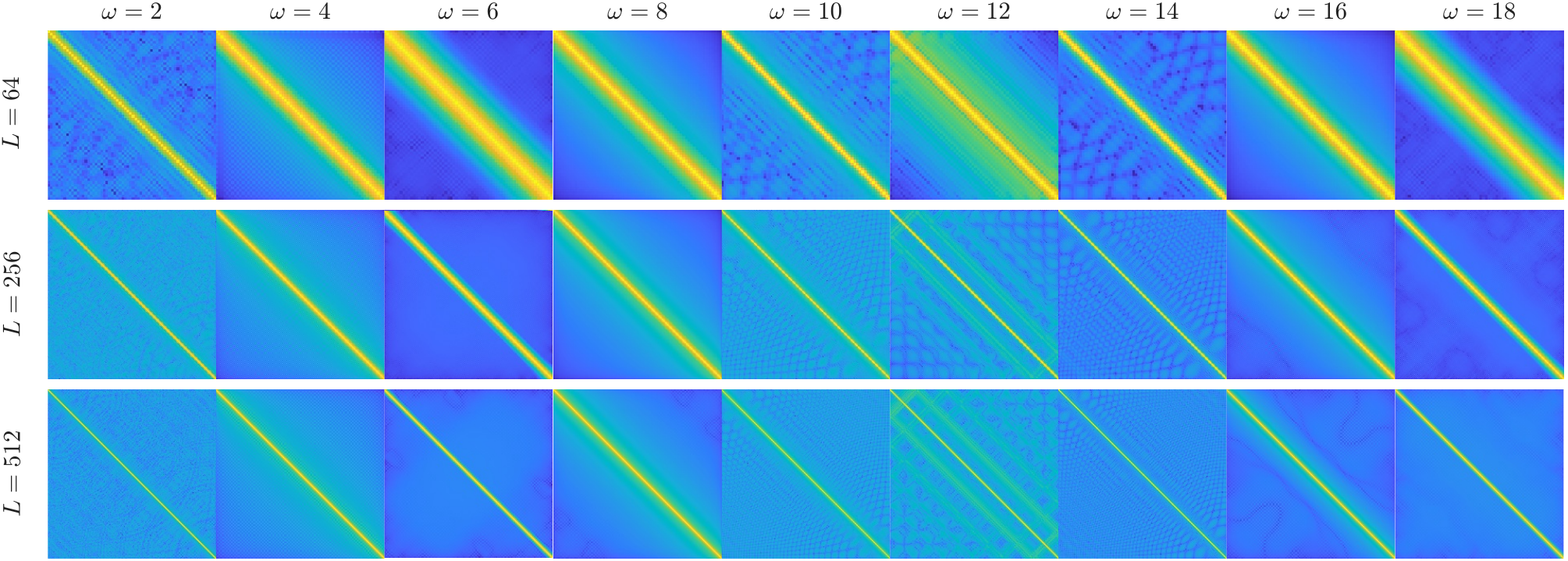}
    \caption{Spatial structure of the steady-state correlation matrix elements for varying system sizes and driving frequencies. The axes correspond to indices $j$ (rows, top-down) and $k$ (columns, left-right). The colors are proportional to $\log |C_{jk}|$, which range from $\log 10^{-16}$ (dark blue) to $\log 1$ (bright yellow). The grid consists of three rows corresponding to system sizes $L = 64, 256, 512$ (top to bottom) and nine columns corresponding to driving frequencies $\omega = 2, 4, 6, 8, 10, 12, 14, 16, 18$ (left to right). The system is driven with amplitude $F=3$, and the static parameters are fixed at $\gamma=0.5$ and $h=3$ (deep in the trivial regime). Resonant driving frequencies (e.g., specific columns) exhibit non-trivial long-range correlations connecting the boundaries, distinct from the short-range behavior observed at off-resonant frequencies.}
    \label{fig:correlation_grid}
\end{figure*}
To demonstrate the effect of the drive, we first examine the spatial structure of the steady-state correlation matrix $C_{jk}^{\uparrow\uparrow}$ within the first chain. (For simplicity, we drop the spin index and use $C_{jk}$ for the rest of the paper.) We fix $h$ deep in the trivial regime ($h > h_c$), where the undriven system ($F=0$) exhibits exponentially decaying short-range correlations. In \cref{fig:correlation_grid}, we illustrate the magnitude of the correlation matrix elements $|C_{jk}|$ at representative points in the parameter space of $L$ and $\omega$. The plot is organized as a grid where rows correspond to increasing system sizes $L \in \{64, 256, 512\}$ and columns correspond to increasing driving frequencies $\omega \in \{2, 4, \dots, 18\}$.

While the off-diagonal elements vanish rapidly for generic driving frequencies, we observe that for specific resonant ranges of $\omega$, significant correlations develop between distant sites. This emergence of long-range order indicates that the boundary drive is capable of bridging the bulk energy modes, effectively mediating correlation transfer across the chain despite the bulk being in a trivially gapped (short-range correlated) phase.

\textit{(b) Phase Diagram and Resonance Mechanism---} To systematically map the regions where this drive-induced phase exists, we construct a phase diagram in the plane of driving frequency $\omega$ and driving strength $F$. We define a global order parameter, the \textit{long-range correlation index} $\chi$ (also called the residual correlator $C_\mathrm{res}$ in Ref.~\cite{PhysRevLett.101.105701}), which measures the average magnitude of correlations between sites separated by more than half the system length:
\begin{equation}
    \chi = \frac{4}{L^2} \sum_{|j-k| > L/2} |C_{jk}|\,.
\end{equation}
This quantity serves as a proxy for the total weight of long-range correlations in the system.

\begin{figure}[t]
    \centering
    \includegraphics[width=\columnwidth]{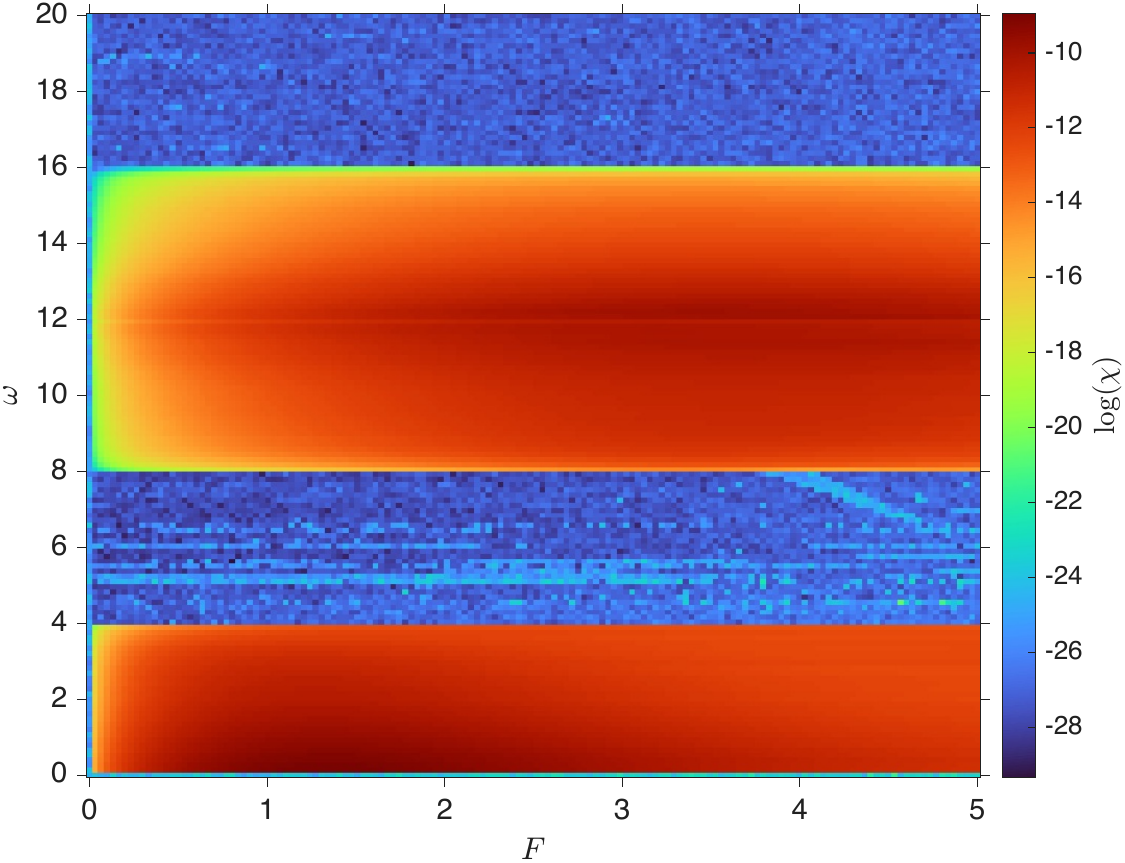}
    \caption{Phase diagram of the long-range correlation index $\chi$ (log scale) as a function of driving strength $F$ and frequency $\omega$. Parameters: $\gamma=0.5, h=3, L=512$. Red regions indicate phases with drive-induced long-range correlations. The distinct domains correspond to specific resonance conditions: $\omega \in [0,4]$ (intra-band) and $\omega \in [8,16]$ (inter-band).}
    \label{fig:phase_diagram}
\end{figure}

\cref{fig:phase_diagram} displays the phase diagram of $\chi$ computed for a system of size $L=512$ with static parameters fixed at $\gamma=0.5$ and $h=3$. We observe distinct \textit{domains} of high correlation separated by regions where correlations remain exponentially suppressed. 

These domains can be understood via a resonance mechanism whereby the driving frequency bridges specific energy gaps in the bulk spectrum \eqref{eq:sepctrum}. For the parameters used here, we identify a low-frequency \textit{intra-band resonance} ($\omega \in [0, 4]$), where the drive energy matches the bandwidth of the individual bands to facilitate internal mixing, and a higher-frequency \textit{inter-band resonance} ($\omega \in [8, 16]$), where the drive bridges the gap between particle and hole branches, creating propagating electron-hole pairs. Readers interested in how these resonant windows shift with different static parameters ($\gamma, h$) are referred to the extensive datasets in \cref{app:moredata}.

Physically, the boundary drive acts as a local source of particles and holes, attempting to inject them coherently into the system. When $\omega$ is off-resonance, the supplied particles and holes have energies unmatched with the energy gap between modes, resulting in induced excitations that are exponentially localized near the boundary ($j=1$). In contrast, when $\omega$ resonates with the bulk dispersion relations, either within a band or bridging the inter-band gap, the boundary excitation propagates and mediates correlation in the bulk, generating the significant long-range correlations captured by the high $\chi$ values in these domains.

\begin{figure}[htbp]
    \centering
    \begin{tabular}[t]{ll}
         (a)& \\
            & \includegraphics[width=0.84\columnwidth]{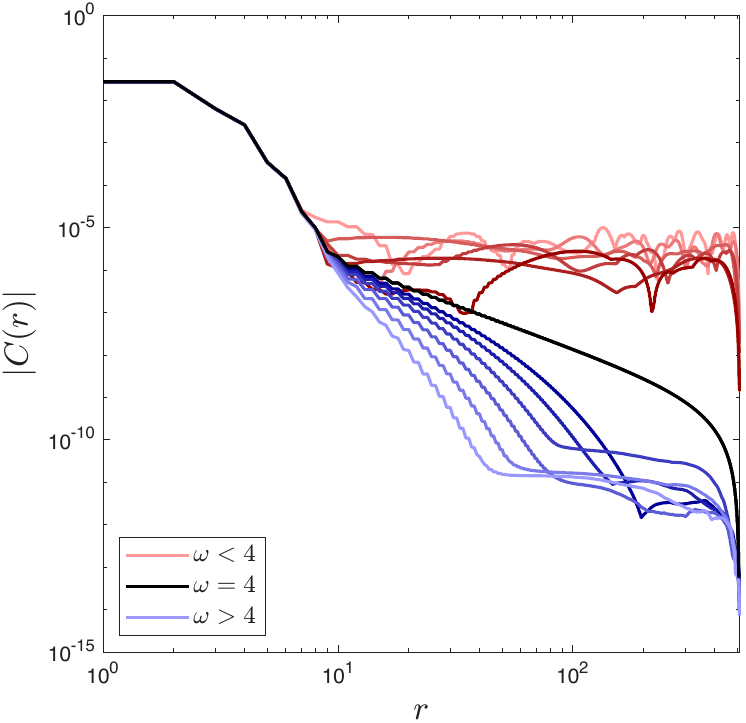} \\
         (b)& \\
         & \includegraphics[width=0.84\columnwidth]{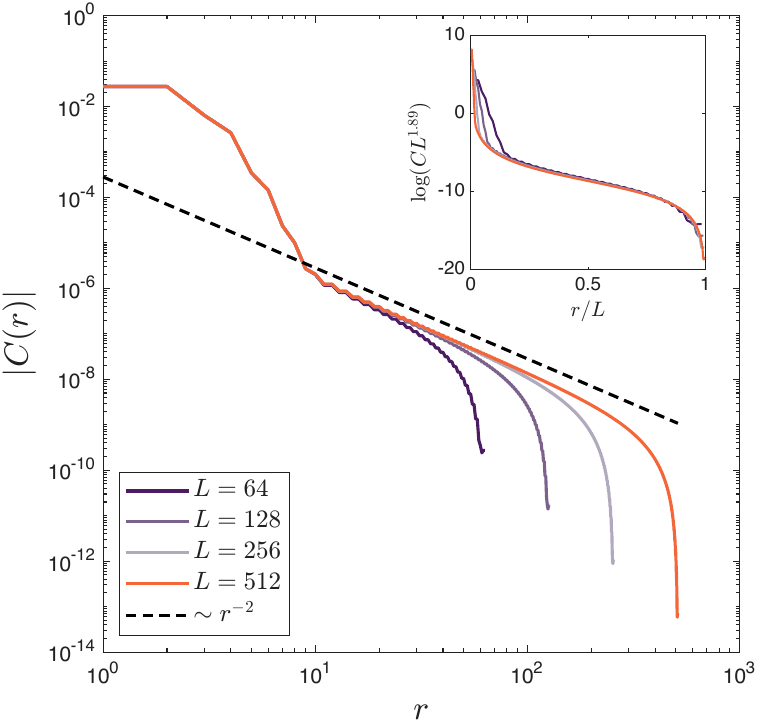} 
    \end{tabular}
    \caption{Detailed analysis of the critical resonance transition at $\omega_c = 4$ ($F=3, \gamma=0.5, h=3$). 
    (a) Spatial profile of correlations $|C(r)|$ for a fixed system size $L=512$ while scanning the driving frequency $\omega$ across the intra-band resonance cutoff. The critical point at $\omega=4$ (black line) exhibits algebraic decay, while slight detuning (red and blue gradients indicate $\omega < 4$ and $\omega > 4$ with $|\Delta\omega| = 2^{-n}$) results in the correlated and trivial phases, respectively.
    (b) Finite-size scaling analysis at the critical frequency $\omega=4$. The main plot shows $|C(r)|$ (with $r=j-k$ and $j+k=L$) for system sizes $L=64, 128, 256, 512$, following a power-law decay $\sim r^{-2}$ (dashed line). The inset demonstrates the data collapse of the scaled correlations $\log(C L^\nu)$ versus $r/L$ (with $\nu=2.0$), confirming the scaling hypothesis.}
    \label{fig:critical_scans}
\end{figure}

To further elucidate the nature of the critical resonance transition, \cref{fig:critical_scans} presents a detailed analysis of the spatial correlations. We probe boundary-to-boundary correlations along the anti-diagonal $j+k=L$. Crucially, at the representative critical frequency $\omega_c = 4$, the correlations exhibit a robust power-law decay, which stands in sharp contrast to the exponential localization observed when the system is detuned from resonance ($\omega >  4$). 

\textit{(c) Scaling with Pairing Potential---} Finally, we investigate the role of the superconducting pairing potential $\gamma$ in facilitating these correlations. The pairing term $\gamma$ in the Hamiltonian is physically responsible for mixing the particle and hole sectors; this mixing is essential for the boundary coupling to effectively inject excitations that can hybridize with the bulk.

To quantify this dependence, we select a representative point within the \textit{inter-band resonance} regime by fixing on-site energy $h=2$, driving frequency $\omega=6$ (within the inter-band resonance regime), and driving strength $F=5$. We then vary $\gamma$ and observe the response of the long-range correlation index $\chi$. (We also examine other choices of parameters $F$ and appropriate $\omega$, and we find the same scaling between $\chi$ and $\gamma$ as stated below.)

\begin{figure}[t]
    \centering
    \includegraphics[width=0.8\columnwidth]{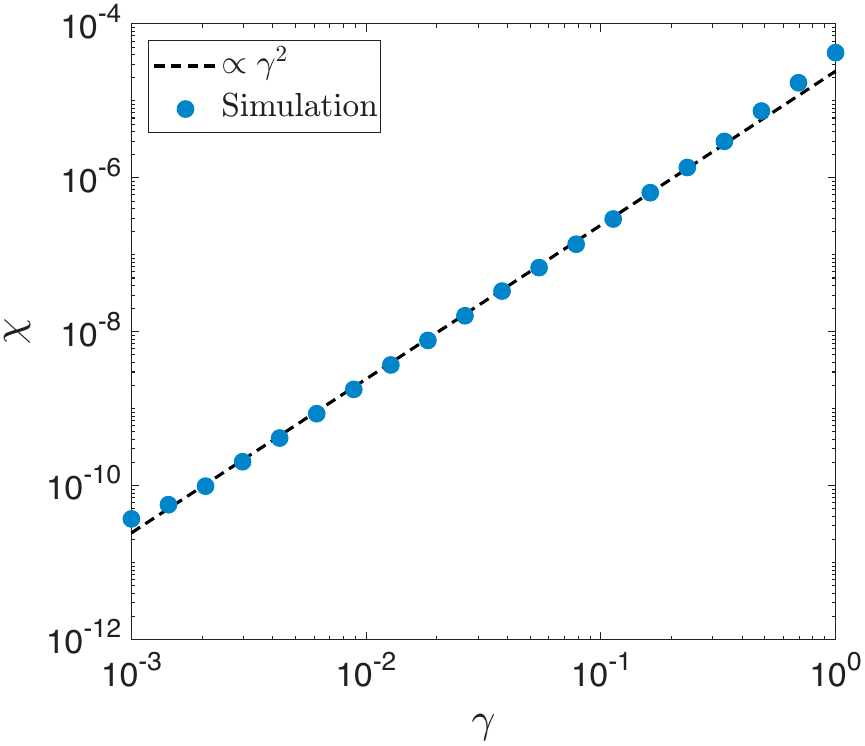}
    \caption{Scaling of the correlation index $\chi$ with the pairing potential $\gamma$ on a log-log scale. The parameters are fixed at $h=2$, $F=5$, and $\omega=6$, placing the system in the inter-band resonance regime. (The system size is $L=512$.) The linear fit indicates a power-law dependence $\chi \propto \gamma^2$, confirming that finite pairing is essential for the drive-induced resonance mechanism.}
    \label{fig:scaling}
\end{figure}

In \cref{fig:scaling}, we plot the correlation index $\chi$ as a function of $\gamma$ on a log-log scale. We observe a clear power-law scaling behavior:
\begin{equation}
    \chi \sim \gamma^2\,.
\end{equation}
The vanishing $\chi$ in the limit $\gamma \to 0$ can be understood physically. In this limit, the Hamiltonian decouples into free fermion bands where particle and hole excitations do not mix. Consequently, the boundary drive, which generates particle-hole pairs, cannot effectively resonate with the bulk Hamiltonian to sustain long-range order without the mediation provided by the pairing term. As shown in \cref{fig:higher_harmonics,fig:phasetransition} of \cref{app:moredata}, we cannot conclusively establish that $\omega=8$ and $\omega=16$ also correspond to genuine critical points.

\paragraph*{Conclusions and Outlook---} In this work, we have investigated the non-equilibrium dynamics of a double fermionic chain subjected to local periodic driving at one end and incoherent dissipation at both boundaries. We have shown that a local boundary drive can induce a transition from a trivial, short-range correlated phase to a non-equilibrium steady state exhibiting long-range order. We attribute this non-equilibrium transition to a resonance between the boundary drive and the bulk spectrum. The drive, which explicitly breaks particle conservation within each chain, acts as a local source of particles and holes. Off resonance, the drive yields only excitations localized near the boundary; whereas, on resonance, matching the bulk dispersion (within a band or across an inter-band gap) produces propagating excitations that transmit correlations through the bulk. This interpretation is supported by the detailed phase diagrams and the analysis of scaling between $\chi$ and $\gamma$, which establishes a quadratic dependence of the long-range correlation index on the pairing potential ($\chi \sim \gamma^2$) when the drive bridges the particle-hole gap. This scaling confirms that particle-hole mixing is the essential ingredient allowing the boundary pairing drive to inject propagating excitations into the bulk.

Our results highlight the utility of coherent boundary control in open quantum systems, suggesting that macroscopic properties can be manipulated without global parameter quenching. This has potential implications for quantum information transfer and state preparation in intermediate-scale quantum devices. Several promising directions exist for future research. First, while our model is non-interacting, extending this analysis to interacting systems (e.g., a Hubbard $U$ term or Heisenberg model) would reveal whether this drive-induced order is robust against, or perhaps enhanced by, interaction. Second, it would be valuable to investigate the interplay between this boundary-drive mechanism and topological phases, particularly to see if the drive can be used to dynamically manipulate Majorana zero modes in the topological phase~\cite{alicea2011non, bauer2019topo}. Third, it will be interesting to extend these studies to long-ranged systems~\cite{purkayastha2021,bijayLR,dhawan2024}, including long-ranged pairing terms, where one could explore the intrinsic interplay between inherent long-ranged versus drive-induced emergent long-ranged correlations. Finally, our protocol is particularly well-suited for implementation in superconducting qubit arrays \cite{mi2022time,mi2022noise} or ultra-cold atoms \cite{weitenberg2011single,nakajima2016topological}, as it aligns naturally with the high degree of local controllability available in these platforms. Introducing a superconducting pairing potential in 1D electron systems via the proximity effect is also an experimentally feasible route and has been established through substantial theoretical \cite{ma1993josephson, Fisher1994, Ostaay2011, Stone2011, Lian2016, Tanaka2021a, kurilovich2023disorder, tam2025quantized} and experimental developments \cite{rickhaus2012quantum,Komatsu2012,amet2016supercurrent, lee2017inducing, park2017propagation, zhao2020interference,  zhao2023loss, hatefipour2022induced}. 

\textit{Acknowledgment ---}We are grateful to Adrian Culver, Haowei Chen, Xi Dai, Yumin Hu, Alex Jacoby, Michael Alexander Kurniawan, Biao Lian, Rahul Roy, Shivaji L. Sondhi, Huaijin Wang, and Frank Zhang for insightful discussions. H.C. receives support from National Science Foundation under award DMR-2141966 and Bede Liu Fund for Excellence at the Department of Electrical and Computer Engineering of Princeton University. W.Z. is supported by the Princeton University Department of Physics. M.K. acknowledges the support of the Department of Atomic Energy, Government of India, under Project No. RTI4001. M.K. thanks the hospitality of the Department of Physics, Princeton University. A.P. thanks KITP for its hospitality during the “Noise-robust Phases of Quantum Matter” program, during which part of this work was completed. This material is based upon work supported by the U.S. Department of Energy, Office of Science, Office of Advanced Scientific Computing Research via the Exploratory Research for Extreme Scale Science (EXPRESS) program under Award Number DE-SC0026216 (A.P.) 

\textit{Disclaimer ---} This report was prepared as an account of work sponsored by an agency of the United States Government. Neither the United States Government nor any agency thereof, nor any of their employees, makes any warranty, express or implied, or assumes any legal liability or responsibility for the accuracy, completeness, or usefulness of any information, apparatus, product, or process disclosed, or represents that its use would not infringe privately owned rights.  Reference herein to any specific commercial product, process, or service by trade name, trademark, manufacturer, or otherwise does not necessarily constitute or imply its endorsement, recommendation, or favoring by the United States Government or any agency thereof. The views and opinions of authors expressed herein do not necessarily state or reflect those of the United States Government or any agency thereof.

\twocolumngrid
\bibliography{library}

\onecolumngrid
\clearpage 
\newpage
\setcounter{equation}{0}
\setcounter{figure}{0}
\renewcommand{\theequation}{S\arabic{equation}}
\renewcommand{\thefigure}{S\arabic{figure}}

\setcounter{section}{0}
\renewcommand{\thesection}{S\arabic{section}}

\section*{\underline{Supplementary Material}}

\section{Rotating Wave}
\label{app:RWA}

\setcounter{equation}{0}
\setcounter{figure}{0}
\renewcommand{\theequation}{A\arabic{equation}}
\renewcommand{\thefigure}{A\arabic{figure}}

Here, we show how we can eliminate the time-dependence of the driving term $H_D(t)$ by performing a unitary transformation to a rotating frame. We define the time-dependent unitary operator:
\begin{equation}
    U(t) = \exp\left( i\frac{\omega\, t}{2} K \right)\,,
\end{equation}
where the Hermitian generator $K$ is defined as:
\begin{equation}
    K = -i\sum_j (c_{j,\uparrow}^\dagger c_{j,\downarrow}^\dagger + c_{j,\uparrow} c_{j,\downarrow})\,.
\end{equation}
This generator satisfies the commutation relations $\left[K, c_{j \uparrow}\right]=ic_{j \downarrow}^{\dagger}, \,\left[K, c_{j \downarrow}\right]=-ic_{j \uparrow}^{\dagger}$, generating an $SO(2)$ rotation between the particle sector of chain $\uparrow$ and the hole sector of chain $\downarrow$.

The density matrix in the rotating frame is $\rho'(t) = U(t)\rho(t)U^\dagger(t)$. The Lindblad equation transforms as:
\begin{equation}
    \frac{d\rho'}{dt} = \mathcal{L}'[\rho'] - i \left[ i \frac{dU}{dt}U^{-1}, \rho' \right].
\end{equation}
Crucially, due to the symmetries of the parameters in \cref{eq:parameter_sym}, the jump operators $L_\mu$ are invariant (up to a phase) or transform into linear combinations such that the dissipative form $\mathcal{D}$ remains time-independent and structurally invariant in the new frame.

The effective Hamiltonian in the rotating frame receives two contributions: the rotated original Hamiltonian and the inertial term (gauge potential) arising from the time-dependent basis:
\begin{equation}
    H_{\text{eff}} = U(t) H(t) U^\dagger(t) + i \frac{dU}{dt} U^\dagger(t)\,.
\end{equation}
Evaluating the unitary rotation, the driving term becomes static:
\begin{equation}
    U(t) H_D(t) U^\dagger(t) = F (N_{1,\uparrow} + N_{1,\downarrow}) + \text{const}\,.
\end{equation}
The inertial term is explicitly:
\begin{equation}
    i \frac{dU}{dt} U^\dagger(t) = - \frac{\omega}{2} K = \frac{\omega}{2} \sum_j (c_{j,\uparrow}^\dagger c_{j,\downarrow}^\dagger + c_{j,\uparrow} c_{j,\downarrow})\,.
\end{equation}
Thus, the total effective Hamiltonian in the rotating frame is time-independent:
\begin{equation}
    H_{\text{eff}} = H_\uparrow + H_\downarrow + F(N_{1,\uparrow} + N_{1,\downarrow}) - \frac{\omega}{2} K\,.
\end{equation}
This allows us to solve for the NESS using standard static methods.

To relate physically observable correlations in the laboratory frame to the computed steady-state correlations in the rotating frame, we perform the inverse transformation. Time-dependent correlation matrix elements are given by:
\begin{equation}
    \braket{c_{j,\alpha}^\dagger c_{k,\beta}}(t) = \Tr[\rho'(t) U(t) c_{j,\alpha}^\dagger c_{k,\beta} U^\dagger(t)]\,,
\end{equation}
where $\alpha, \beta \in \{\uparrow, \downarrow\}$. In the NESS, $\rho'(t) \to \rho_{\text{SS}}'$ is time-independent. The unitary conjugation of the operators yields the following Bogoliubov transformations (using $\theta = \omega t/2$):
\begin{subequations}
\begin{align}
    U(t) c_{j,\uparrow}^\dagger U^\dagger(t) &= \cos\theta \, c_{j,\uparrow}^\dagger + \sin\theta \, c_{j,\downarrow}\,, \\
    U(t) c_{j,\downarrow} U^\dagger(t) &= \cos\theta \, c_{j,\downarrow} - \sin\theta \, c_{j,\uparrow}^\dagger\,.
\end{align}
\end{subequations}

We are interested in the time-averaged correlations $\overline{\braket{\mathcal{O}}} = \frac{1}{T}\int_0^T dt \braket{\mathcal{O}}(t)$. Upon averaging, oscillating terms ($\sin\theta \cos\theta$) vanish, while $\overline{\cos^2\theta} = \overline{\sin^2\theta} = \frac{1}{2}$.

For the intra-chain correlations ($\uparrow\uparrow$):
\begin{equation}
\begin{aligned}
    \overline{\braket{c_{j,\uparrow}^\dagger c_{k,\uparrow}}} &= \frac{1}{2} \braket{c_{j,\uparrow}^\dagger c_{k,\uparrow}}_R + \frac{1}{2} \braket{c_{j,\downarrow} c_{k,\downarrow}^\dagger}_R \\
    &= \frac{1}{2} \braket{c_{j,\uparrow}^\dagger c_{k,\uparrow}}_R + \frac{1}{2} (\delta_{jk} - \braket{c_{k,\downarrow}^\dagger c_{j,\downarrow}}_R)\,.
\end{aligned}
\end{equation}
where in the rotating frame
\begin{equation}
    \langle \mathcal{O} \rangle_R = \mathrm{Tr}\left[\rho_{\text{SS}}\mathcal{O}\right].
\end{equation}
Similarly, for the $\downarrow\downarrow$ chain, noting that the hole particle transformation introduces a sign change in the definition of the hole density:
\begin{equation}
    \overline{\braket{c_{j,\downarrow}^\dagger c_{k,\downarrow}}} = \frac{1}{2}\left(\braket{ c_{j,\downarrow}^\dag c_{k,\downarrow}}_R - \braket{ c_{k,\uparrow}^\dag c_{j,\uparrow}}_R + \delta_{jk}\right).
\end{equation}

For the inter-chain mixing correlations ($\uparrow\downarrow$):
\begin{equation}
    \overline{\braket{c_{j,\uparrow}^\dag c_{k,\downarrow}}} = \frac{1}{2}\braket{ c_{j,\uparrow}^\dag c_{k,\downarrow}}_R - \frac{1}{2}\braket{ c_{k,\uparrow}^\dag c_{j,\downarrow}}_R.
\end{equation}

\section{Supporting Numerical Data}
\label{app:moredata}

\setcounter{equation}{0}
\setcounter{figure}{0}
\renewcommand{\theequation}{B\arabic{equation}}
\renewcommand{\thefigure}{B\arabic{figure}}

Here, we provide a comprehensive numerical survey of the phase diagram to elucidate how the resonance structure evolves across the phase transition ($h_c = 1-\gamma^2$) and with varying coupling strengths.

\begin{figure}
    \centering
    \includegraphics[width=\textwidth]{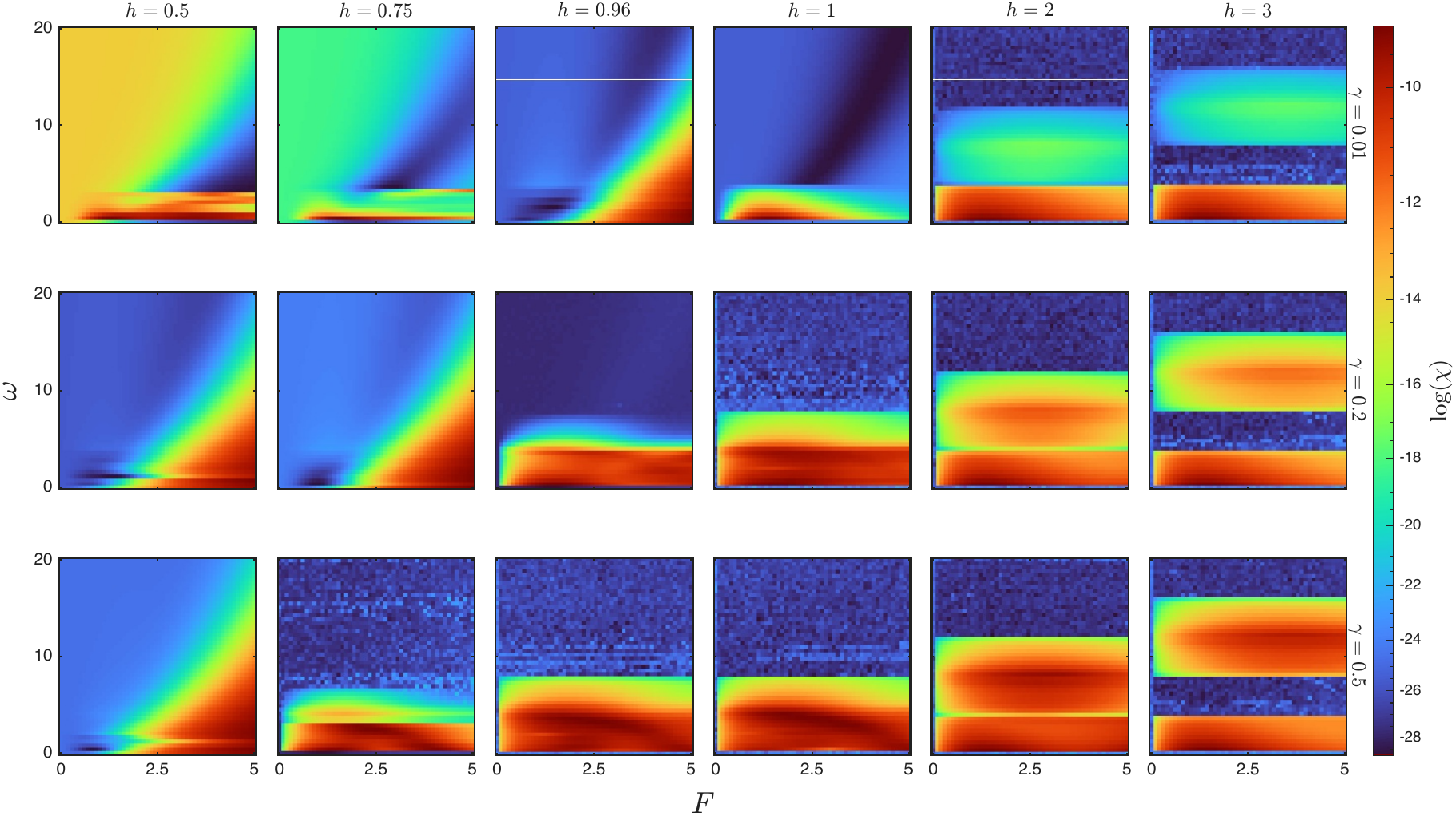}
    \caption{Comprehensive scan of the $F$-$\omega$ phase diagram for the long-range correlation index $\chi$. The grid is arranged by pairing potential \textbf{(Rows: $\gamma = 0.01, 0.2, 0.5$)} and on-site energy \textbf{(Columns: $h = 0.5, 0.75, 0.96, 1, 2, 3$)}. (The system size is $L=512$.)}
    \label{fig:appendix_grid}
\end{figure}

\cref{fig:appendix_grid} displays a grid of phase diagrams for the long-range correlation index $\chi$ in the $(F, \omega)$ plane. The rows correspond to increasing pairing potentials $\gamma \in \{0.01, 0.2, 0.5\}$ (top to bottom), while the columns correspond to increasing on-site energy $h \in \{0.5, 0.75, 0.96, 1, 2, 3\}$ (left to right), spanning from the deep correlated phase through the critical point to the deep trivial phase. (The system size is $L=512$.)

Two key observations can be drawn from this extensive dataset:

\begin{itemize}
    \item In the correlated region $h< 1 - \gamma^2$, the resonance mechanism differs qualitatively from the trivial case discussed in the main text. We observe that the regions of high correlation depend strongly on the driving strength $F$, suggesting that the drive is not merely bridging a bulk gap but is likely interacting non-trivially with the Majorana edge modes characteristic of this phase.

    \item In the trivial region $h> 1 - \gamma^2$, we can clearly verify the scaling discussed in the main text. Comparing the phase diagrams from the bottom row ($\gamma = 0.5$) to the top row ($\gamma = 0.01$), the high-correlation domains associated with inter-band resonances rapidly fade and eventually vanish. This visual evidence corroborates that as $\gamma \to 0$, the mechanism for inter-band mixing is lost, and $\chi$ consequently drops to zero.
\end{itemize}

\begin{figure}
    \centering
    \includegraphics[width=0.83\textwidth]{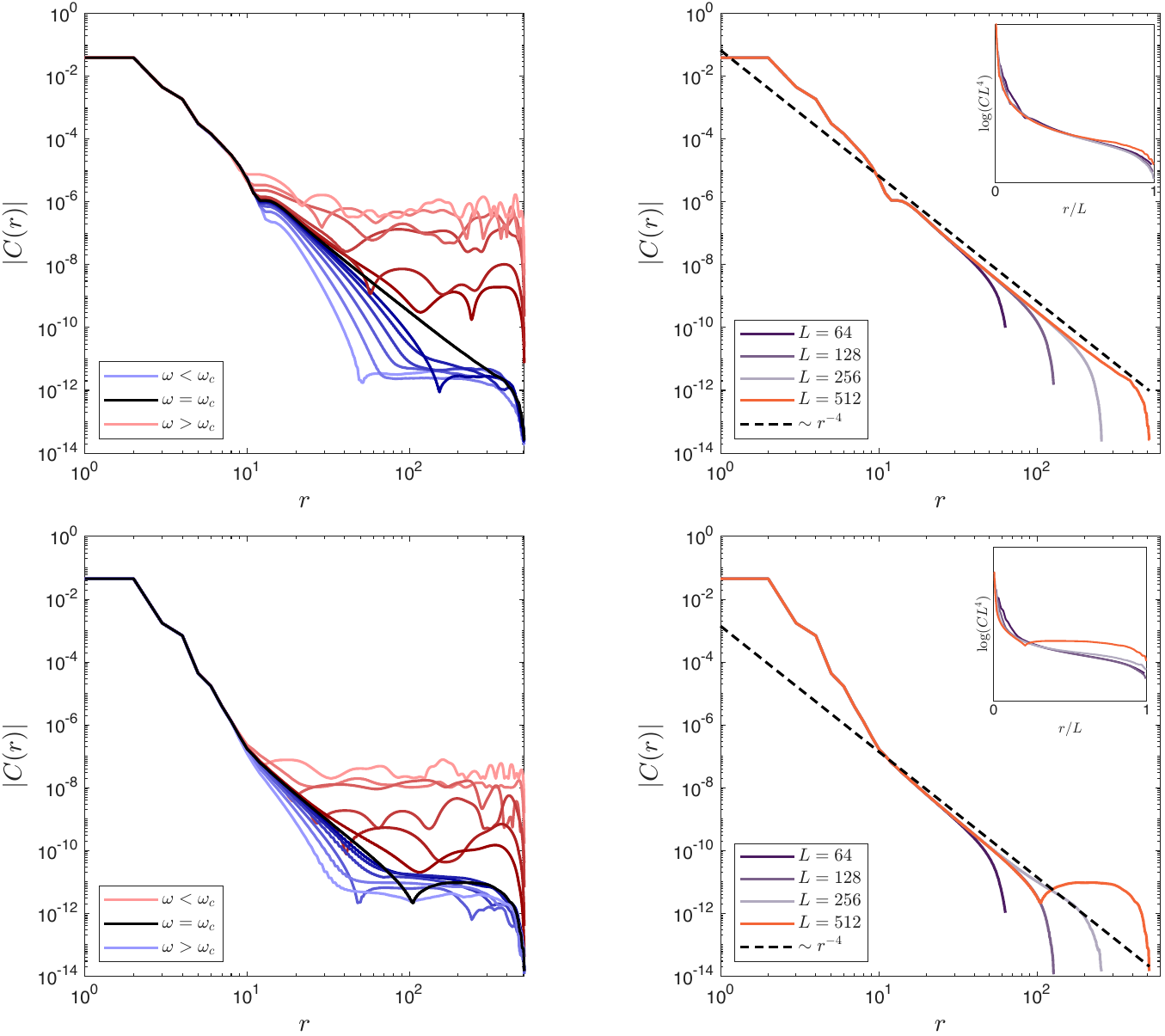}   
    \caption{Universality of the critical transition at higher harmonic resonances ($\omega_c = 8, 16$, $F=3, \gamma=0.5, h=3, L=512$).
    Upper panels: Analysis of the $\omega=8$ resonance. Upper Left: Frequency scan showing the correlated phase (red), critical point (black), and trivial phase (blue). Note the color inversion for $\omega=8$ (red for $\omega > \omega_c$). Upper Right: Scaling collapse with $\nu=4$.
    Lower panels: Analysis of the $\omega=16$ resonance. Lower Left: Frequency scan. Lower Right: Scaling collapse consistent with $\nu=4$.}
    \label{fig:higher_harmonics}
\end{figure}

\begin{figure}
    \centering
    \includegraphics[width=0.78\textwidth]{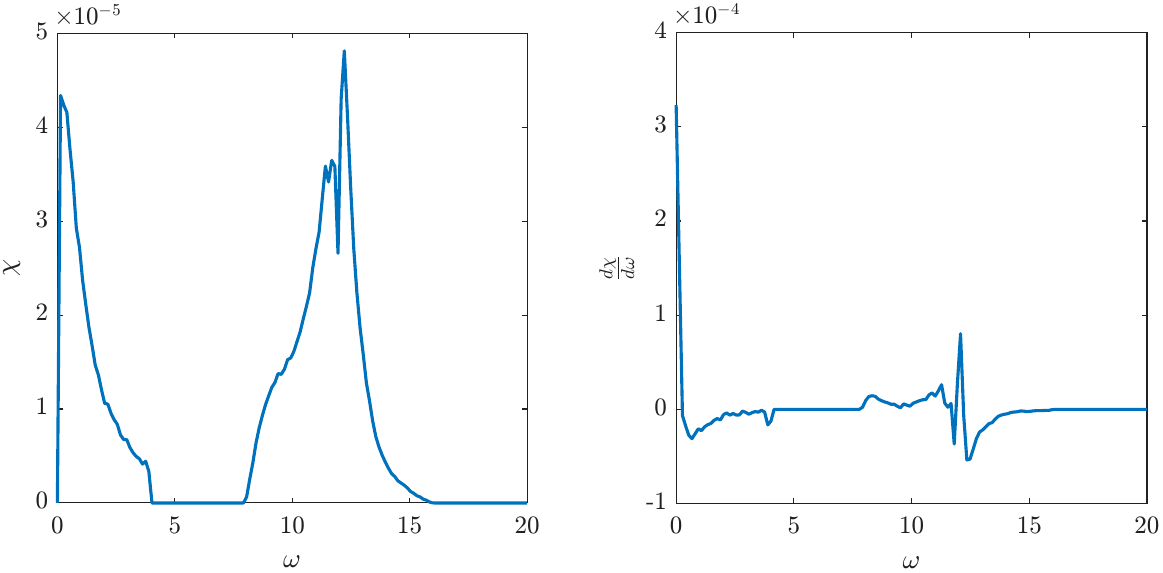}
    \caption{Long-range correlation index $\chi$ and its frequency derivative. The left panel shows the correlation index $\chi$ as a function of frequency $\omega$ for the parameter $F \approx 3.0$. The right panel displays the derivative $d\chi/d\omega$. (The system size is $L=512$.)}
    \label{fig:phasetransition}
\end{figure}

We also extend our analysis to higher frequencies $\omega$. \Cref{fig:higher_harmonics} displays the steady-state correlation profiles for $\omega \approx 8$ and $\omega \approx 16$ ($F=3, \gamma=0.5, h=3$). Similar to the primary resonance at $\omega=4$, we observe a distinct algebraic decay $|C(r)| \sim r^{-\nu}$ exactly at the critical drive frequencies, while detuning leads to the order and correlated phases. The scaling collapse (insets) yields a consistent critical exponent $\nu \approx 4$ across all observed resonances, suggesting a unified universality class for the transition. \Cref{fig:phasetransition} quantifies how long-range correlations change as the boundary-drive frequency is tuned at $F \approx 3$. As shown in \cref{fig:higher_harmonics,fig:phasetransition}, it is inconclusive from these results whether $\omega=8$ and $\omega=16$ are critical points.

\end{document}

%% file: fig1.tex
\tikzset{every picture/.style={line width=0.75pt}} 

\begin{tikzpicture}[x=0.75pt,y=0.75pt,yscale=-0.8,xscale=0.8]

\draw  [color={rgb, 255:red, 155; green, 155; blue, 155 }  ,draw opacity=1 ][fill={rgb, 255:red, 244; green, 178; blue, 188 }  ,fill opacity=1 ] (66,68.83) .. controls (66,60.83) and (72.49,54.33) .. (80.5,54.33) .. controls (88.51,54.33) and (95,60.83) .. (95,68.83) .. controls (95,76.84) and (88.51,83.33) .. (80.5,83.33) .. controls (72.49,83.33) and (66,76.84) .. (66,68.83) -- cycle ;
\draw  [color={rgb, 255:red, 155; green, 155; blue, 155 }  ,draw opacity=1 ][fill={rgb, 255:red, 244; green, 178; blue, 188 }  ,fill opacity=1 ] (126.33,68.83) .. controls (126.33,60.83) and (132.83,54.33) .. (140.83,54.33) .. controls (148.84,54.33) and (155.33,60.83) .. (155.33,68.83) .. controls (155.33,76.84) and (148.84,83.33) .. (140.83,83.33) .. controls (132.83,83.33) and (126.33,76.84) .. (126.33,68.83) -- cycle ;
\draw    (95,67.33) -- (126.33,67.33)(95,70.33) -- (126.33,70.33) ;
\draw  [color={rgb, 255:red, 155; green, 155; blue, 155 }  ,draw opacity=1 ][fill={rgb, 255:red, 244; green, 178; blue, 188 }  ,fill opacity=1 ] (186.67,68.83) .. controls (186.67,60.83) and (193.16,54.33) .. (201.17,54.33) .. controls (209.17,54.33) and (215.67,60.83) .. (215.67,68.83) .. controls (215.67,76.84) and (209.17,83.33) .. (201.17,83.33) .. controls (193.16,83.33) and (186.67,76.84) .. (186.67,68.83) -- cycle ;
\draw    (155.33,67.33) -- (186.67,67.33)(155.33,70.33) -- (186.67,70.33) ;
\draw  [color={rgb, 255:red, 155; green, 155; blue, 155 }  ,draw opacity=1 ][fill={rgb, 255:red, 244; green, 178; blue, 188 }  ,fill opacity=1 ] (278.67,69.17) .. controls (278.67,61.16) and (285.16,54.67) .. (293.17,54.67) .. controls (301.17,54.67) and (307.67,61.16) .. (307.67,69.17) .. controls (307.67,77.17) and (301.17,83.67) .. (293.17,83.67) .. controls (285.16,83.67) and (278.67,77.17) .. (278.67,69.17) -- cycle ;
\draw    (247.33,67.67) -- (278.67,67.67)(247.33,70.67) -- (278.67,70.67) ;
\draw  [color={rgb, 255:red, 155; green, 155; blue, 155 }  ,draw opacity=1 ][fill={rgb, 255:red, 165; green, 200; blue, 239 }  ,fill opacity=1 ] (68.4,126.83) .. controls (68.4,118.83) and (74.89,112.33) .. (82.9,112.33) .. controls (90.91,112.33) and (97.4,118.83) .. (97.4,126.83) .. controls (97.4,134.84) and (90.91,141.33) .. (82.9,141.33) .. controls (74.89,141.33) and (68.4,134.84) .. (68.4,126.83) -- cycle ;
\draw  [color={rgb, 255:red, 155; green, 155; blue, 155 }  ,draw opacity=1 ][fill={rgb, 255:red, 165; green, 200; blue, 239 }  ,fill opacity=1 ] (128.73,126.83) .. controls (128.73,118.83) and (135.23,112.33) .. (143.23,112.33) .. controls (151.24,112.33) and (157.73,118.83) .. (157.73,126.83) .. controls (157.73,134.84) and (151.24,141.33) .. (143.23,141.33) .. controls (135.23,141.33) and (128.73,134.84) .. (128.73,126.83) -- cycle ;
\draw    (97.4,125.33) -- (128.73,125.33)(97.4,128.33) -- (128.73,128.33) ;
\draw  [color={rgb, 255:red, 155; green, 155; blue, 155 }  ,draw opacity=1 ][fill={rgb, 255:red, 165; green, 200; blue, 239 }  ,fill opacity=1 ] (189.07,126.83) .. controls (189.07,118.83) and (195.56,112.33) .. (203.57,112.33) .. controls (211.57,112.33) and (218.07,118.83) .. (218.07,126.83) .. controls (218.07,134.84) and (211.57,141.33) .. (203.57,141.33) .. controls (195.56,141.33) and (189.07,134.84) .. (189.07,126.83) -- cycle ;
\draw    (157.73,125.33) -- (189.07,125.33)(157.73,128.33) -- (189.07,128.33) ;
\draw  [color={rgb, 255:red, 155; green, 155; blue, 155 }  ,draw opacity=1 ][fill={rgb, 255:red, 165; green, 200; blue, 239 }  ,fill opacity=1 ] (281.07,127.17) .. controls (281.07,119.16) and (287.56,112.67) .. (295.57,112.67) .. controls (303.57,112.67) and (310.07,119.16) .. (310.07,127.17) .. controls (310.07,135.17) and (303.57,141.67) .. (295.57,141.67) .. controls (287.56,141.67) and (281.07,135.17) .. (281.07,127.17) -- cycle ;
\draw    (249.73,125.67) -- (281.07,125.67)(249.73,128.67) -- (281.07,128.67) ;
\draw [color={rgb, 255:red, 208; green, 2; blue, 27 }  ,draw opacity=1 ]   (82.9,112.33) .. controls (75.75,107.87) and (73.43,93.36) .. (78.62,85.55) ;
\draw [shift={(80.5,83.33)}, rotate = 137.52] [fill={rgb, 255:red, 208; green, 2; blue, 27 }  ,fill opacity=1 ][line width=0.08]  [draw opacity=0] (5.36,-2.57) -- (0,0) -- (5.36,2.57) -- cycle    ;
\draw [color={rgb, 255:red, 208; green, 2; blue, 27 }  ,draw opacity=1 ]   (80.5,83.33) .. controls (87.33,86.76) and (90.22,98.17) .. (84.27,109.89) ;
\draw [shift={(82.9,112.33)}, rotate = 301.56] [fill={rgb, 255:red, 208; green, 2; blue, 27 }  ,fill opacity=1 ][line width=0.08]  [draw opacity=0] (5.36,-2.57) -- (0,0) -- (5.36,2.57) -- cycle    ;
\draw [color={rgb, 255:red, 128; green, 128; blue, 128 }  ,draw opacity=1 ]   (66,68.83) .. controls (63.79,69.64) and (62.28,68.93) .. (61.47,66.72) .. controls (60.66,64.51) and (59.15,63.8) .. (56.94,64.61) .. controls (54.73,65.42) and (53.21,64.71) .. (52.4,62.5) .. controls (51.59,60.29) and (50.08,59.58) .. (47.87,60.39) .. controls (45.66,61.2) and (44.15,60.49) .. (43.34,58.28) -- (40.6,57) -- (40.6,57) ;
\draw [shift={(50.94,61.82)}, rotate = 24.98] [fill={rgb, 255:red, 128; green, 128; blue, 128 }  ,fill opacity=1 ][line width=0.08]  [draw opacity=0] (7.14,-3.43) -- (0,0) -- (7.14,3.43) -- (4.74,0) -- cycle    ;
\draw [color={rgb, 255:red, 128; green, 128; blue, 128 }  ,draw opacity=1 ]   (42.2,79.8) .. controls (43.01,77.59) and (44.53,76.89) .. (46.74,77.71) .. controls (48.95,78.53) and (50.47,77.83) .. (51.28,75.62) .. controls (52.09,73.41) and (53.61,72.71) .. (55.82,73.52) .. controls (58.03,74.34) and (59.55,73.64) .. (60.36,71.43) .. controls (61.18,69.22) and (62.7,68.52) .. (64.91,69.34) -- (66,68.83) -- (66,68.83) ;
\draw [shift={(55.37,73.73)}, rotate = 155.26] [fill={rgb, 255:red, 128; green, 128; blue, 128 }  ,fill opacity=1 ][line width=0.08]  [draw opacity=0] (5.36,-2.57) -- (0,0) -- (5.36,2.57) -- (3.56,0) -- cycle    ;
\draw [color={rgb, 255:red, 128; green, 128; blue, 128 }  ,draw opacity=1 ]   (67.6,126.83) .. controls (65.39,127.64) and (63.88,126.93) .. (63.07,124.72) .. controls (62.26,122.51) and (60.75,121.8) .. (58.54,122.61) .. controls (56.33,123.42) and (54.81,122.71) .. (54,120.5) .. controls (53.19,118.29) and (51.68,117.58) .. (49.47,118.39) .. controls (47.26,119.2) and (45.75,118.49) .. (44.94,116.28) -- (42.2,115) -- (42.2,115) ;
\draw [shift={(52.54,119.82)}, rotate = 24.98] [fill={rgb, 255:red, 128; green, 128; blue, 128 }  ,fill opacity=1 ][line width=0.08]  [draw opacity=0] (7.14,-3.43) -- (0,0) -- (7.14,3.43) -- (4.74,0) -- cycle    ;
\draw [color={rgb, 255:red, 128; green, 128; blue, 128 }  ,draw opacity=1 ]   (43.8,137.8) .. controls (44.61,135.59) and (46.13,134.89) .. (48.34,135.71) .. controls (50.55,136.53) and (52.07,135.83) .. (52.88,133.62) .. controls (53.69,131.41) and (55.21,130.71) .. (57.42,131.52) .. controls (59.63,132.34) and (61.15,131.64) .. (61.96,129.43) .. controls (62.78,127.22) and (64.3,126.52) .. (66.51,127.34) -- (67.6,126.83) -- (67.6,126.83) ;
\draw [shift={(56.97,131.73)}, rotate = 155.26] [fill={rgb, 255:red, 128; green, 128; blue, 128 }  ,fill opacity=1 ][line width=0.08]  [draw opacity=0] (5.36,-2.57) -- (0,0) -- (5.36,2.57) -- (3.56,0) -- cycle    ;
\draw [color={rgb, 255:red, 128; green, 128; blue, 128 }  ,draw opacity=1 ]   (308.07,69.76) .. controls (308.88,67.54) and (310.39,66.83) .. (312.6,67.64) .. controls (314.81,68.45) and (316.33,67.74) .. (317.14,65.53) .. controls (317.95,63.32) and (319.46,62.61) .. (321.67,63.42) .. controls (323.88,64.23) and (325.39,63.52) .. (326.2,61.31) .. controls (327.01,59.1) and (328.52,58.39) .. (330.73,59.2) -- (334.6,57.4) -- (334.6,57.4) ;
\draw [shift={(323.69,62.48)}, rotate = 155.02] [fill={rgb, 255:red, 128; green, 128; blue, 128 }  ,fill opacity=1 ][line width=0.08]  [draw opacity=0] (7.14,-3.43) -- (0,0) -- (7.14,3.43) -- (4.74,0) -- cycle    ;
\draw [color={rgb, 255:red, 128; green, 128; blue, 128 }  ,draw opacity=1 ]   (332.93,81.21) .. controls (330.72,82.03) and (329.2,81.33) .. (328.39,79.12) .. controls (327.58,76.91) and (326.06,76.21) .. (323.85,77.03) .. controls (321.64,77.84) and (320.12,77.14) .. (319.31,74.93) .. controls (318.49,72.72) and (316.97,72.02) .. (314.76,72.84) .. controls (312.55,73.66) and (311.03,72.96) .. (310.22,70.75) -- (308.07,69.76) -- (308.07,69.76) ;
\draw [shift={(319.23,74.9)}, rotate = 24.74] [fill={rgb, 255:red, 128; green, 128; blue, 128 }  ,fill opacity=1 ][line width=0.08]  [draw opacity=0] (5.36,-2.57) -- (0,0) -- (5.36,2.57) -- (3.56,0) -- cycle    ;
\draw [color={rgb, 255:red, 128; green, 128; blue, 128 }  ,draw opacity=1 ]   (311.2,128.33) .. controls (312.01,126.12) and (313.52,125.41) .. (315.73,126.22) .. controls (317.94,127.03) and (319.45,126.32) .. (320.26,124.11) .. controls (321.07,121.9) and (322.59,121.19) .. (324.8,122) .. controls (327.01,122.81) and (328.52,122.1) .. (329.33,119.89) .. controls (330.14,117.68) and (331.65,116.97) .. (333.86,117.78) -- (337.73,115.97) -- (337.73,115.97) ;
\draw [shift={(326.82,121.06)}, rotate = 155.02] [fill={rgb, 255:red, 128; green, 128; blue, 128 }  ,fill opacity=1 ][line width=0.08]  [draw opacity=0] (7.14,-3.43) -- (0,0) -- (7.14,3.43) -- (4.74,0) -- cycle    ;
\draw [color={rgb, 255:red, 128; green, 128; blue, 128 }  ,draw opacity=1 ]   (336.06,139.79) .. controls (333.85,140.61) and (332.33,139.91) .. (331.52,137.7) .. controls (330.71,135.49) and (329.19,134.79) .. (326.98,135.6) .. controls (324.77,136.42) and (323.25,135.72) .. (322.43,133.51) .. controls (321.62,131.3) and (320.1,130.6) .. (317.89,131.42) .. controls (315.68,132.24) and (314.16,131.54) .. (313.35,129.33) -- (311.2,128.33) -- (311.2,128.33) ;
\draw [shift={(322.36,133.48)}, rotate = 24.74] [fill={rgb, 255:red, 128; green, 128; blue, 128 }  ,fill opacity=1 ][line width=0.08]  [draw opacity=0] (5.36,-2.57) -- (0,0) -- (5.36,2.57) -- (3.56,0) -- cycle    ;

\draw (216,59.2) node [anchor=north west][inner sep=0.75pt]    {$\cdots $};
\draw (97.2,52.8) node [anchor=north west][inner sep=0.75pt]  [font=\scriptsize]  {$t_{\uparrow } ,\gamma _{\uparrow }$};
\draw (158,52.4) node [anchor=north west][inner sep=0.75pt]  [font=\scriptsize]  {$t_{\uparrow } ,\gamma _{\uparrow }$};
\draw (249.2,53.2) node [anchor=north west][inner sep=0.75pt]  [font=\scriptsize]  {$t_{\uparrow } ,\gamma _{\uparrow }$};
\draw (75.2,40) node [anchor=north west][inner sep=0.75pt]  [font=\scriptsize]  {$h_{\uparrow }$};
\draw (133.6,40.4) node [anchor=north west][inner sep=0.75pt]  [font=\scriptsize]  {$h_{\uparrow }$};
\draw (194,40) node [anchor=north west][inner sep=0.75pt]  [font=\scriptsize]  {$h_{\uparrow }$};
\draw (286.4,40) node [anchor=north west][inner sep=0.75pt]  [font=\scriptsize]  {$h_{\uparrow }$};
\draw (218.4,117.2) node [anchor=north west][inner sep=0.75pt]    {$\cdots $};
\draw (99.4,130.23) node [anchor=north west][inner sep=0.75pt]  [font=\scriptsize]  {$t_{\downarrow } ,\gamma _{\downarrow }$};
\draw (76.8,143.2) node [anchor=north west][inner sep=0.75pt]  [font=\scriptsize]  {$h_{\downarrow }$};
\draw (159.73,130.23) node [anchor=north west][inner sep=0.75pt]  [font=\scriptsize]  {$t_{\downarrow } ,\gamma _{\downarrow }$};
\draw (251.73,130.57) node [anchor=north west][inner sep=0.75pt]  [font=\scriptsize]  {$t_{\downarrow } ,\gamma _{\downarrow }$};
\draw (137.2,143.6) node [anchor=north west][inner sep=0.75pt]  [font=\scriptsize]  {$h_{\downarrow }$};
\draw (198.4,144) node [anchor=north west][inner sep=0.75pt]  [font=\scriptsize]  {$h_{\downarrow }$};
\draw (290,144) node [anchor=north west][inner sep=0.75pt]  [font=\scriptsize]  {$h_{\downarrow }$};
\draw (74.8,59.6) node [anchor=north west][inner sep=0.75pt]   [align=left] {1};
\draw (77.6,118.4) node [anchor=north west][inner sep=0.75pt]   [align=left] {1};
\draw (136,59.6) node [anchor=north west][inner sep=0.75pt]   [align=left] {2};
\draw (138.4,118.4) node [anchor=north west][inner sep=0.75pt]   [align=left] {2};
\draw (196.4,60.4) node [anchor=north west][inner sep=0.75pt]   [align=left] {3};
\draw (198.4,117.6) node [anchor=north west][inner sep=0.75pt]   [align=left] {3};
\draw (90,92.4) node [anchor=north west][inner sep=0.75pt]  [font=\scriptsize,color={rgb, 255:red, 208; green, 2; blue, 27 }  ,opacity=1 ]  {$F,\omega $};
\draw (287.2,60.4) node [anchor=north west][inner sep=0.75pt]    {$L$};
\draw (290.4,118.8) node [anchor=north west][inner sep=0.75pt]    {$L$};
\draw (6,49.8) node [anchor=north west][inner sep=0.75pt]  [font=\scriptsize,color={rgb, 255:red, 128; green, 128; blue, 128 }  ,opacity=1 ]  {$\Gamma _{1,\uparrow ,l}$};
\draw (6,71) node [anchor=north west][inner sep=0.75pt]  [font=\scriptsize,color={rgb, 255:red, 128; green, 128; blue, 128 }  ,opacity=1 ]  {$\Gamma _{1,\uparrow ,g}$};
\draw (337.6,52.2) node [anchor=north west][inner sep=0.75pt]  [font=\scriptsize,color={rgb, 255:red, 128; green, 128; blue, 128 }  ,opacity=1 ]  {$\Gamma _{L,\uparrow ,l}$};
\draw (338,73.4) node [anchor=north west][inner sep=0.75pt]  [font=\scriptsize,color={rgb, 255:red, 128; green, 128; blue, 128 }  ,opacity=1 ]  {$\Gamma _{L,\uparrow ,g}$};
\draw (340,111.4) node [anchor=north west][inner sep=0.75pt]  [font=\scriptsize,color={rgb, 255:red, 128; green, 128; blue, 128 }  ,opacity=1 ]  {$\Gamma _{L,\downarrow ,l}$};
\draw (340.4,132.6) node [anchor=north west][inner sep=0.75pt]  [font=\scriptsize,color={rgb, 255:red, 128; green, 128; blue, 128 }  ,opacity=1 ]  {$\Gamma _{L,\downarrow ,g}$};
\draw (6,108.6) node [anchor=north west][inner sep=0.75pt]  [font=\scriptsize,color={rgb, 255:red, 128; green, 128; blue, 128 }  ,opacity=1 ]  {$\Gamma _{1,\downarrow ,l}$};
\draw (6,129.8) node [anchor=north west][inner sep=0.75pt]  [font=\scriptsize,color={rgb, 255:red, 128; green, 128; blue, 128 }  ,opacity=1 ]  {$\Gamma _{1,\downarrow ,g}$};

\end{tikzpicture}